\documentclass[
aip,
reprint,
amsmath,
amssymb,
hyphens
]{revtex4-1}
\usepackage{graphicx} 
\usepackage{dcolumn} 
\usepackage{bm}  
\usepackage[detect-all]{siunitx} 
\usepackage{doi}  
\usepackage{url}  
\providecommand{\urlprefix}{URL: } 
\hypersetup{colorlinks, linkcolor=blue, urlcolor=blue}  
\usepackage{xfrac}  
\usepackage{enumitem}  
\usepackage{pgfplots}  
\pgfplotsset{compat=newest}  
\usepackage[utf8]{inputenc}  
\DeclareUnicodeCharacter{2212}{-}
\usepackage{ctable}  
\usepackage{multirow}  
\newcolumntype{C}{>{\centering\arraybackslash}X}  
\newcolumntype{d}[1]{D{,}{\; \rightarrow \;}{#1}} 
\usepackage{adjustbox}  
\usepackage{stackengine}  
\usepackage{chngcntr}  

\DeclareMathOperator{\E}{E}
\DeclareMathOperator{\Var}{Var}

\begin{document}

\title{Wide-band Parametric Amplifier Readout and Resolution of Optical Microwave Kinetic Inductance Detectors}

\author{Nicholas Zobrist}
\email[]{nzobrist@physics.ucsb.edu}
\affiliation{\mbox{Department of Physics, University of California, Santa Barbara, CA 93106, USA}}

\author{Byeong Ho Eom}
\author{Peter Day}
\affiliation{\mbox{Jet Propulsion Laboratory, California Institute of Technology, Pasadena, California 91125, USA}}

\author{Benjamin A. Mazin}
\homepage{http://www.mazinlab.org}
\affiliation{\mbox{Department of Physics, University of California, Santa Barbara, CA 93106, USA}}

\author{Seth R. Meeker}
\author{Bruce Bumble}
\author{Henry G. LeDuc}
\affiliation{\mbox{Jet Propulsion Laboratory, California Institute of Technology, Pasadena, California 91125, USA}}

\author{Grégoire Coiffard}
\affiliation{\mbox{Department of Physics, University of California, Santa Barbara, CA 93106, USA}}

\author{Paul Szypryt}
\affiliation{National Institute of Standards and Technology, Boulder, Colorado 80305, USA}

\author{Neelay Fruitwala}
\author{Isabel Lipartito}
\author{Clint Bockstiegel}
\affiliation{\mbox{Department of Physics, University of California, Santa Barbara, CA 93106, USA}}

\date{\today}
\begin{abstract}
The energy resolution of a single photon counting Microwave Kinetic Inductance Detector (MKID) can be degraded by noise coming from the primary low temperature amplifier in the detector's readout system. Until recently, quantum limited amplifiers have been incompatible with these detectors due to dynamic range, power, and bandwidth constraints. However, we show that a kinetic inductance based traveling wave parametric amplifier can be used for this application and reaches the quantum limit. The total system noise for this readout scheme was equal to \num{\sim2.1} in units of quanta. For incident photons in the \SIrange{800}{1300}{nm} range, the amplifier increased the average resolving power of the detector from \numrange{\sim6.7}{9.3} at which point the resolution becomes limited by noise on the pulse height of the signal. Noise measurements suggest that a resolving power of up to \num{25} is possible if redesigned detectors can remove this additional noise source.
\end{abstract}
\maketitle
\onecolumngrid
Optical MKIDs~\cite{Szypryt2017b} are superconducting, single photon counting, energy resolving sensors which are sensitive to radiation in the ultraviolet to near infrared range. Advantages over semiconductor detectors in this wavelength band include the absence of false counts (read noise, dark current, and cosmic rays), intrinsic spectral resolution, high speed, and radiation hardness. Other superconducting detectors have shown promise at these wavelengths,~\cite{Martin2006, Burney2006} but they are difficult to chain together into large arrays. Optical MKIDs, however, are naturally frequency domain multiplexed, which has enabled full-scale instruments at the Palomar observatory~\cite{Mazin2013, Meeker2018} and Subaru telescope.~\cite{Walter2018} In the future, these detectors will be included in a balloon borne mission.~\cite{Cook2015}

Photon counting MKIDs operate differently than MKIDs designed for longer wavelength detection in the bolometric regime. Instead of measuring a constant flux of photons, they record individual photon events similarly to an X-ray calorimeter. To achieve a measurable detector response for a single photon event they tend to be smaller and able to handle less signal power than their longer wavelength bolometric counterparts. In these conditions, amplifier noise can be comparable in magnitude to the detector phase noise that originates from microscopic two-level system (TLS) states on the surface or between layers of the device.~\cite{Gao2008c} The TLS noise can be mitigated through careful sample preparation,~\cite{DeGraaf2018} fabrication,~\cite{Megrant2012, Calusine2018} and device design~\cite{Noroozian2009, Mignot2018, Vissers2012} while the effect of amplifier noise can be addressed by designing detectors that can handle higher signal powers.~\cite{Zobrist2018, Beldi2019} These routes are actively pursued, but, for optical MKIDs, improving the main readout amplifier's noise floor offers an additional path to lowering the total system noise.

Quantum mechanics imposes an uncertainty relationship between the two quadratures of an electromagnetic signal.~\cite{Caves1982} This relationship results in a lower limit to the noise that a high gain, phase-preserving, linear amplifier adds to its input signal, equal to that of the electromagnetic zero-point fluctuations $\left( A=\sfrac{1}{2} \right)$. To readout an optical MKID array, the primary amplifier must have a moderately high saturation power, significant dynamic range, and a large bandwidth. High electron mobility transistor (HEMT) amplifiers are often used to satisfy these requirements.~\cite{Wadefalk2003, Schleeh2013} State of the art commercial HEMT amplifiers operating over \SIrange{4}{8}{GHz} at \SI{5}{K} typically reach added noise numbers, in units of quanta, as low as $A = \num{8.0}$ (\SI{2.3}{K} noise temperature at \SI{6}{GHz}).~\cite{Schleeh2012, LNF} A quantum limited amplifier that could operate in these conditions would reduce the added amplifier noise by a factor of \num{16}.

Parametric amplifiers have been shown to perform at or near to the quantum limit and have been used in several experiments with superconducting resonators. A \SI{\sim 1}{MHz}, bandwidth format was used to investigate a superconducting resonator's noise properties in the dissipation quadrature.~\cite{Gao2011} However, only one resonator could be measured at a time, and a carrier suppression tone was needed, which made the amplifier incompatible with single photon measurements. At low signal powers, larger bandwidth formats have been able to read out up to \num{20} superconducting qubits.~\cite{Macklin2015} This readout system, too, does not have enough dynamic range for an optical MKID array as the amplifier's saturation power is on the order of the required signal tone's power.

Traveling wave parametric amplifiers (TWPA) based on a superconductor's nonlinear kinetic inductance were designed to handle this wide-band, high power and dynamic range case.~\cite{Zmuidzinas2014, Chaudhuri2017, Eom2012} In this paper we show how one of these parametric amplifiers can be integrated with a large MKID array and demonstrate its ability to measure single photon events with quantum limited amplifier noise.

\begin{figure}[t]
\def\svgwidth{\linewidth}
\trimbox{3pt 3pt 3pt 3pt}{\input{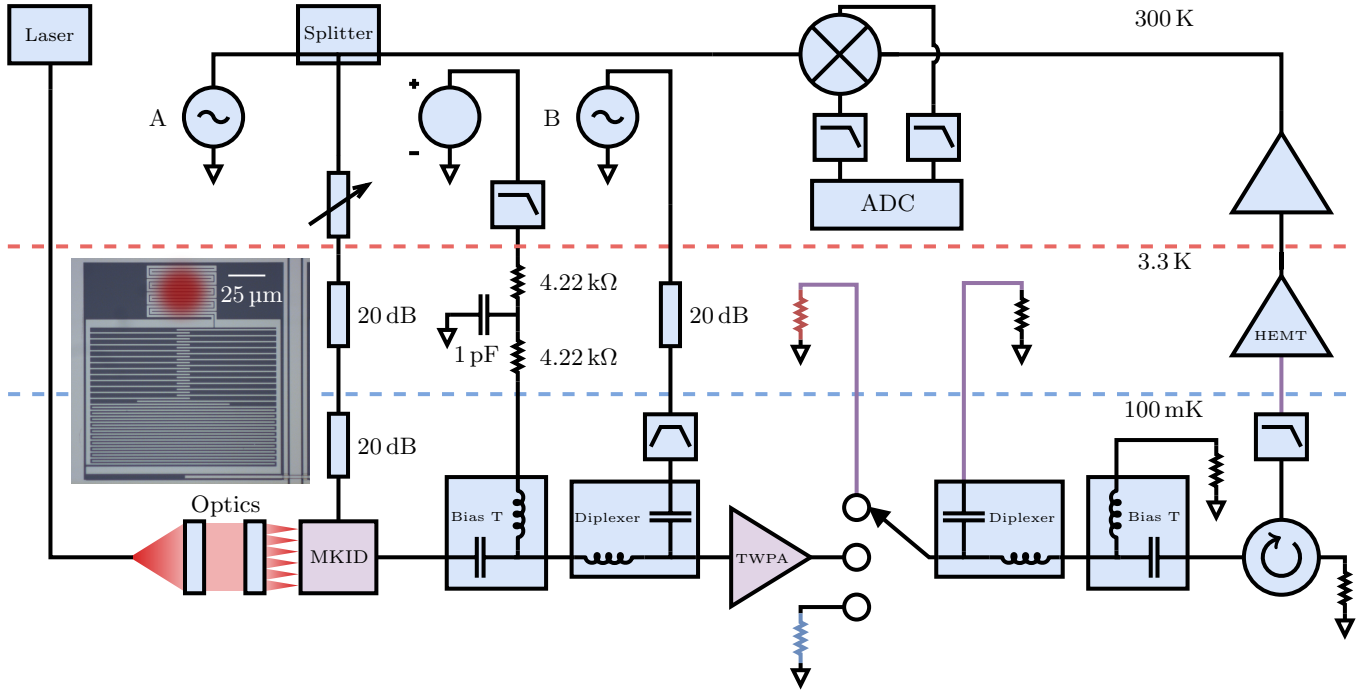}}
\caption{Circuit diagram of the readout system. The two synthesizers are labeled A and B for the signal and pump tones respectively. \SI{50}{\ohm} terminations are represented as resistors to ground, and superconducting components are shaded purple. The diagram does not include line losses. The system noise is measured using a cryogenic switch which can connect the HEMT amplifier, to the parametric amplifier, or to one of two matched \SI{50}{\ohm} loads on different temperature stages (shaded red and blue). Light from a laser is directed to the MKID array with an optical fiber and focused onto each inductor with a collimating lens and microlens array. The approximate spot size on one resonator in the array is shown in red in the inset image.} \label{fig:block}
\end{figure}

The full readout is divided between three temperature stages with most of the large electronics at room temperature. The rest of the components are cooled to either \SI{3.3}{K} or \SI{100}{mK} in a Leiden Cryogenics CF-200 dilution refrigerator with the MKID array and parametric amplifier inside of a Amumetal 4K magnetic shield developed by Amuneal.  Outside of the cryostat we employ a homodyne readout system with the signals digitized at a sample rate of \SI{2}{MHz} after being low pass filtered at \SI{1}{MHz} to prevent aliasing. Figure~\ref{fig:block} shows a schematic of the setup used for this experiment.

The detector that we tested was designed for the MEC instrument at the Subaru telescope on Mauna Kea in Hawaii.~\cite{Walter2018} It has ten niobium coplanar waveguide feedlines and a \num{20440} pixel platinum silicide MKID array multiplexed over \SIrange{4}{8}{GHz}.~\cite{Szypryt2017b} Each pixel is a lumped element resonator, capacitively coupled to one of the feedlines. A microscope image of a pixel is shown in the inset of figure~\ref{fig:block}.

For the photon measurements, we used five single mode laser diodes with wavelengths from \SIrange{808}{1310}{nm}. The \SI{1120}{nm} diode was purchased from Eagle Yard, and the others were purchased from Thor Labs. The diode light was coupled into a fiber using an integrating sphere and sent into the fridge. Near the detector, the fiber output was collimated and directed toward a microlens array on top of the device. The microlens array focused the light onto the inductor of each pixel and was purchased from Advanced Microoptic Systems. It is made from 1 mm thick STIH53 glass with \num{140 x 146} lenslets at a \SI{150}{\micro m} pitch.

We chose a resonator from the array at \SI{5.67446}{GHz} with a coupling and internal quality factors of \num{15100} and \num{190000} respectively for this test. All data taken off resonance was captured at a frequency of \SI{5.675}{GHz}. Near the resonance frequency the parametric amplifier had a constant gain of \SI{13.7}{dB}, and the resonator was driven almost to saturation at about \SI{-106}{dBm}. When the parametric amplifier is turned on, we see changes in the coupling quality factor of some resonators on the order of \SI{\sim 10}{\percent}. More isolation between the parametric amplifier and MKID array could remove this effect, but since this difference is smaller than the variation intrinsic to the design, modifications were unnecessary for this test.

The parametric amplifier itself is a wide-band, traveling-wave, kinetic inductance amplifier of the general type first described by Eom et al.,~\cite{Eom2012} but is an updated version and differs in several respects. The coplanar wave guide transmission line structure uses finer features (e.g. the center line width and gaps are \SI{320}{nm}). In order to lower the characteristic impedance to \SI{50}{\ohm}, added capacitance is provided with a interdigitated structure in a similar manner to the amplifier described by Chaudhuri et al.~\cite{Chaudhuri2017} Additionally, the amplifier is operated in a three-wave mixing mode by applying a DC bias current using the technique shown by Vissers et al.~\cite{Vissers2016} Further details about this amplifier are in preparation and will be published separately.

Operation of the amplifier using three-wave mixing requires the use of both a pump tone and a DC bias current.  The pump tone (from synthesizer B in figure~\ref{fig:block}) at \SI{14.765}{GHz} is attenuated by 20 dB at \SI{3.3}{K} and filtered using a \SIrange{14.5}{17}{GHz} Marki FB-1575 bandpass filter to ensure that phase noise from the pump generator does not leak into the signal band. The pump tone power at the input of the amplifier is about \SI{-23}{dBm}. The tone is then combined with a DC current of approximately \SI{0.7}{mA} to produce about \SI{15}{dB} of gain from \SIrange{5}{10}{GHz}. The two Anritsu K250 bias tees and the two Marki DPX-1114 diplexers isolate the DC current and pump tone from the other components, respectively. The diplexers have over \SI{60}{dB} of isolation at the pump frequency, and the pump tone is terminated on the cryostat ground at \SI{3.3}{K} where there is more cooling power. The parametric amplifier is then protected from reflections off of the warmer components with a \SIrange{3}{11}{GHz} Pamtech CTH1365K10 isolator.

\bgroup
\def\arraystretch{1.5}
\ctable[
caption={Maximum likelihood estimates for the input, parametric amplifier, HEMT, and total system noise in units of quanta at \SI{5.675}{GHz} (from left to right). $1\sigma$ statistical errors then systematic errors are reported to the right of the estimates. The HEMT noise level corresponds to a noise temperature of about \SI{5.3}{K} and is within the manufacturer's specifications for the amplifier. The effective noise temperature of the combined parametric and HEMT amplifiers is \SI{0.43}{K}. The measurement used to determine these numbers and errors is detailed in appendices~\ref{sec:noise_calc} and~\ref{sec:systematics}.},
label={tab:noise},
width=\columnwidth,
pos=t,
botcap
]{
C C C C
}{
}{
\toprule[0.08em]  
$A_\text{I}$ & $A_\text{P}$ & $A_\text{H}$ & $A_\text{sys}$ \\
\cmidrule[0.03em](r{.5em}){1-3} \cmidrule[0.03em](l{.5em}){4-4}
0.71\stackanchor{\tiny +0.03}{\tiny -0.03} \stackanchor{\tiny +0.2}{\tiny -0.01} &
0.59\stackanchor{\tiny +0.02}{\tiny -0.02} \stackanchor{\tiny +0.2}{\tiny -0.07} &
19.5\stackanchor{\tiny +0.1}{\tiny -0.1} \stackanchor{\tiny +2}{\tiny -2} &
2.13\stackanchor{\tiny +0.01}{\tiny -0.01} \stackanchor{\tiny +0.8}{\tiny -0.1} \\
\bottomrule[0.08em]
}
\egroup

Input signal saturation occurs when the amplified signal power at the output of the amplifier is approximately \SI{15}{dB} below the pump power. At that point, the pump amplitude becomes depleted and the operating point of the amplifier is altered. For \SI{15}{dB} gain, this results in an input signal saturation power of around \SI{-53}{dBm}. The large saturation power obviates the need for any carrier suppression tones. For this array, even if all of the signal tone power from a simultaneous measurement of a full feedline of resonators reached the amplifier input, it would take \SI{-86}{dBm} of power per tone before reaching saturation, well above the typical operating point of an optical MKID. In practice, most of the signal tone power is reflected by the MKIDs before reaching the parametric amplifier, so higher signal powers might be usable.

The second low temperature amplification stage is a \SIrange{4}{12}{GHz} CIT412 HEMT amplifier from Cosmic Microwave Technology with about \SI{32}{dB} of gain and thermalized to the \SI{3.3}{K} temperature stage with a copper heat strap. A low pass filter is included before the amplifier input to ensure that the pump tone does not leak past the diplexers and saturate the HEMT. This amplifier is required because the parametric amplifier does not have enough gain to boost the signal above a standard room temperature amplifier's noise floor.

Since the transmission through the parametric amplifier is near unity when unpowered, we can perform an extended Y-factor measurement of the system. This method allows us to accurately determine the noise components and assign them to different elements in the setup. Table~\ref{tab:noise} shows this breakdown for a signal tone slightly detuned from the MKID resonance frequency. The details of the procedure used to collect this data are laid out in appendices~\ref{sec:noise_calc} and~\ref{sec:systematics}.

While the parametric amplifier is performing near optimally, some aspects of the system may be improved.
The observed system noise is about two times the achievable limit. Almost \SI{75}{\percent} of this excess comes from the HEMT amplifier. This effect could be mitigated by either increasing the parametric amplifier's gain or using a lower noise secondary amplifier. Additionally, the input noise is about \SI{0.2}{quanta} larger than one would expect from a \SI{100}{mK} termination. It could be brought closer to its lower bound by increasing the lowest temperature attenuation to block more of the thermal noise from the \SI{3.3}{K} and \SI{300}{K} stages.

\begin{figure}[t]
\resizebox{0.5\columnwidth}{!}{\trimbox{14pt 17pt 15pt 13pt}{\input{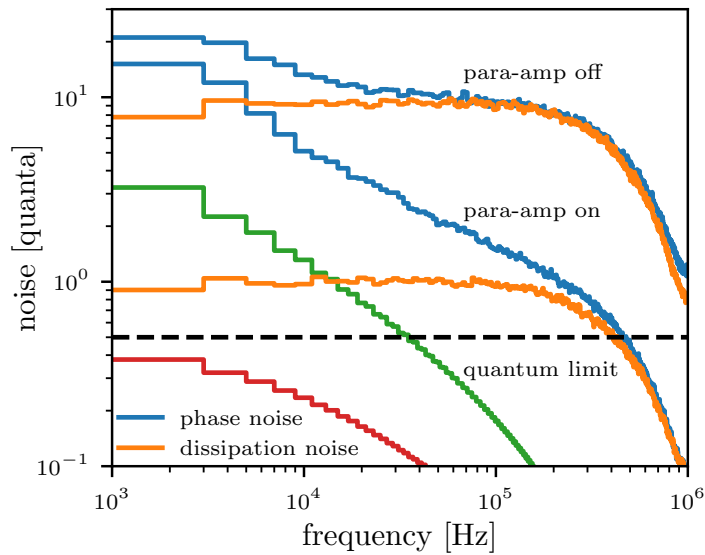}}}
\caption{Noise on resonance in both signal quadratures with the parametric amplifier's pump tone and DC current on and off. When on, the white noise level decreases to about a factor of two above the quantum limit of a half photon per quadrature (zero-point plus amplifier noise). For an average single photon event, the magnitudes of the Fourier transforms, in arbitrary units, are shown for the phase (green) and dissipation (red) quadratures. The majority of the useful signal lies between \num{2} and \SI{60}{kHz}.} \label{fig:noise}
\end{figure}

Figure~\ref{fig:noise} shows the noise in the bandwidth of our signal with the signal tone set to the MKID resonance frequency. This state is used for all of the single photon measurements. The flat component of the noise decreased by a factor of 9.6 when the parametric amplifier was powered, with the noise level in the dissipation response quadrature matching the off resonance noise floor for the system. In the phase response quadrature, there is significant low frequency noise which we attribute to TLS noise in the detector.

When the detector is illuminated, we measure a two dimensional pulse record for each photon event. From a record, we can calculate a maximum likelihood estimate of the photon's energy and arrival time. More details of this estimation are given in appendices~\ref{sec:response} and~\ref{sec:fitting}. Since the pulse decay time is about \SI{35}{\micro s}, care was taken to use low count rates (\SI{<200}{Hz}) and to exclude photons which arrived within \SI{2}{ms} of each other to ensure clean event records.

\begin{figure}[t]
\resizebox{0.5\columnwidth}{!}{\trimbox{16pt 15pt 15pt 13pt}{\input{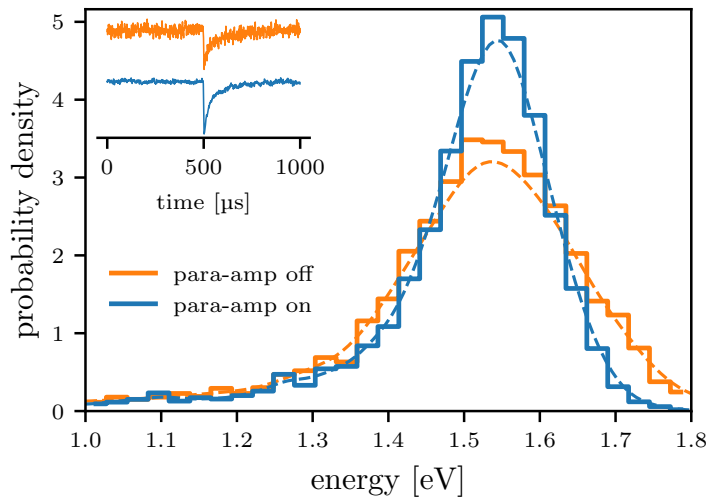}}}
\caption{The normalized distribution of fitted energies is plotted when the device was illuminated with a \SI{808}{nm} laser. There is a clear reduction of the line width when the parametric amplifier's DC current and pump tone are turned on. However, the resulting line shape is significantly skewed to lower energies. The dashed lines are kernel density estimations used to calculate the resolving powers listed in table~\ref{tab:R}. The inset shows the relative improvement in data quality by comparing an arbitrary phase response in the two conditions.} \label{fig:resolution}
\end{figure}

The resolving power of the detector, $\sfrac{E}{\Delta E}$, is determined by measuring many photon events of known energy and evaluating the full-width half-max (FWHM) of the resulting fitted-energy distribution. Figure~\ref{fig:resolution} shows this distribution for \SI{808}{nm} photons with the parametric amplifier's pump tone and DC current both on and off, along with the kernel density estimation of the distribution that was used to compute the FWHM. We note that even though the dissipation noise is significantly smaller than the phase noise when the parametric amplifier is on, the smaller dissipation signal for this detector means that roughly \SI{90}{\percent} of the expected resolving power can be achieved with the phase signal alone.

Using the parametric amplifier clearly improves the resolving power, but the results do not match the expected values for the noise level in the system. Table~\ref{tab:R} shows this discrepancy over the five measured photon energies. Our noise model does not account for the small, non-stationary decrease in phase noise during a photon event, but we still see this discrepancy when estimating the energy using only the dissipation signal, which has stationary noise. Additionally, since the actual phase noise for a given event is smaller than a stationary noise model predicts, non-stationary noise is unlikely to explain the measured discrepancy in resolving power.

The resolving power is approximately the inverse of a scaled standard deviation, so we can consider additional noise on the pulse height, corresponding to a resolving power of $R_0$, using
\begin{equation} \label{eq:R0}
    \frac{1}{R_\text{measured}^2} = \frac{1}{R_0^2} + \frac{1}{R_\text{expected}^2},
\end{equation}
where $R_\text{measured}$ and $R_\text{expected}$ are the resolving powers from the columns in table~\ref{tab:R}. Since $R_\text{expected}$ represents a lower bound, $R_0$ calculated from equation~\ref{eq:R0} can be interpreted as a upper bound. For this data, $R_0$ ranges from \numrange{7.3}{11} with the parametric amplifier off and from \numrange{9.6}{10} with the amplifier on. These results are in rough agreement with a constant detector-related energy uncertainty of \num{\sim10}.

$R_0$ may be attributable to more than one source. In appendix~\ref{sec:current}, we detail how the non-uniform current density in the resonators can contribute to an uncertainty in pulse height and account for the skewed distribution in figure~\ref{fig:resolution}. This effect is strongly dependent on the diffusion constant for platinum silicide, which is unknown for our films. However, the diffusion constants of similar films~\cite{Baturina2005} suggest $R_0 \sim \numrange{20}{40}$.  Pulse shape variations may also contribute to $R_0$, but we expect them to be small since the quasiparticle distribution averages out the current non-uniformities at time scales on the order of the quasiparticle recombination rate. In our data, we do not see significant variations in the pulse shape for photons of the same energy with different pulse heights.

Interactions between the quasiparticles and phonons in the superconductor could also introduce fluctuations in the pulse height. The standard Fano-limit~\cite{Fano1947} is well above the measured resolving powers, but hot phonon escape from the superconductor to the substrate during the initial energy down-conversion may be an important factor.~\cite{Kozorezov2007, Kozorezov2008} For an \SI{800}{nm} photon absorbed in platinum silicide on a sapphire substrate, this effect could contribute anywhere from $R_0 \sim\numrange{13}{30}$---with the uncertainty being dominated by the unknown electron-phonon interaction energy in platinum silicide. See appendix~\ref{sec:phonon} for more details on this calculation.

Together, the worst case estimates of $R_0$ for the current non-uniformity and the hot phonon escape effects may account for the measured value of $R_0 \sim 10$. More detailed measurements of the superconducting properties of platinum silicide are needed to make a more accurate estimation. Nevertheless, the lowered amplifier noise reveals that the detector's design is the limiting factor in its resolving power. With the help of a parametric amplifier, future designs addressing this issue could reach resolving powers of up to \num{25} without requiring lower TLS noise.

In summary, we have demonstrated a straightforward readout scheme for a full MKID array which incorporates a quantum limited amplifier and presented measurements showing that this amplifier is compatible with single photon detection. Detailed calculations of the noise in this system show that it significantly outperforms the standard HEMT amplifier readout scheme at the cost of a modest increase in complexity and exposes an additional noise source on the pulse height for these detectors.

N.Z. was supported throughout this work by a NASA Space Technology Research Fellowship. The MKID arrays used were developed under NASA grant NNX16AE98G. This research was carried out in part at the Jet Propulsion Laboratory and California Institute of Technology, under a contract with the National Aeronautics and Space Administration.

\bgroup
\def\arraystretch{1.5}
\ctable[
caption={The measured increase in the resolving power is shown by comparing data with the parametric amplifier unpowered and powered for different photon energies. A lower bound on the expected resolving power of the estimate, from equation~\ref{eq:variance}, is also specified (see appendix~\ref{sec:fitting}). The detector response starts to saturate, as designed, at the highest energies, so the resolving power begins to decrease. Results are shown for data reduced using both the phase and dissipation, just the phase, and just the dissipation signals.},
label={tab:R},
pos=b,
width=\columnwidth,
botcap
]{
@{}Xd{3.3}d{3.2}d{3.3}d{3.2}d{3.3}d{3.3}
}{
}{
\toprule[0.08em]  
\multirow{3}*{\centering Energy [\si{eV}]} & \multicolumn{6}{c}{Resolving Power $\left[\sfrac{E}{\Delta E}\right]$}  \\
\cmidrule[0.03em](l{.5em}){2-7}
\multicolumn{1}{c}{} & \multicolumn{2}{c}{Phase and Dissipation} & \multicolumn{2}{c}{Phase} & \multicolumn{2}{c}{Dissipation} \\
\cmidrule[0.03em](l{.5em}r{.5em}){2-3} \cmidrule[0.03em](l{.5em}r{.5em}){4-5} \cmidrule[0.03em](l{.5em}){6-7}
\multicolumn{1}{c}{} & \multicolumn{1}{C}{Measured} & \multicolumn{1}{C}{Expected}  & \multicolumn{1}{C}{Measured} & \multicolumn{1}{C}{Expected}  & \multicolumn{1}{C}{Measured} & \multicolumn{1}{C}{Expected} \\
\cmidrule[0.03em](r{.5em}){1-1} \cmidrule[0.03em](l{.5em}r{.5em}){2-2}  \cmidrule[0.03em](l{.5em}r{.5em}){3-3} \cmidrule[0.03em](l{.5em}r{.5em}){4-4} \cmidrule[0.03em](l{.5em}r{.5em}){5-5} \cmidrule[0.03em](l{.5em}r{.5em}){6-6} \cmidrule[0.03em](l{.5em}){7-7}
\num{1.53} (\SI{808}{nm}) & \num{5.8} , \num{8.9}  & \num{9.5} , \num{23} & \num{5.4} , \num{8.8} & \num{8.7} , \num{22} & \num{1.4} , \num{5.9} & \num{1.8} , \num{8.7} \\
\num{1.35} (\SI{920}{nm}) & \num{7.4} , \num{9.4} & \num{10} , \num{24} & \num{6.6} , \num{9.1} & \num{9.7} , \num{23} & \num{1.4} , \num{5.8} & \num{1.7} , \num{7.8} \\
\num{1.27} (\SI{980}{nm}) & \num{7.5} , \num{9.6} & \num{11} , \num{25} & \num{6.6} , \num{9.3} & \num{10} , \num{23} & \num{1.5} , \num{6.8} & \num{1.6} , \num{8.8} \\
\num{1.11} (\SI{1120}{nm}) & \num{6.6} , \num{9.6} & \num{9.3} , \num{24} & \num{6.1} , \num{9.2} & \num{8.8} , \num{22} & \num{1.9} , \num{6.9} & \num{1.8} , \num{9.9} \\
\num{0.946} (\SI{1310}{nm}) & \num{6.0} , \num{9.2} & \num{8.7} , \num{23} & \num{5.5} , \num{8.9} & \num{8.3} , \num{20} & \num{1.9} , \num{6.1} & \num{1.9} , \num{9.9} \\
\bottomrule[0.08em]
}
\egroup
\appendix
\counterwithout{equation}{section}
\renewcommand{\theequation}{S\arabic{equation}}
\renewcommand{\thefigure}{S\arabic{figure}}
\renewcommand{\thetable}{S\Roman{table}}
\setcounter{equation}{0}
\setcounter{figure}{0}
\setcounter{table}{0}
\section{Noise Measurement} \label{sec:noise_calc}
The noise figures presented in table~\ref{tab:noise} are determined by the following measurement, which consists of taking four noise data sets using a microwave switch to change the input to the HEMT. From each noise data set, the power spectrum is computed for each quadrature and summed. Superconducting coaxes are used between the HEMT and the switch and between the switch and the terminations to ensure an accurate calibration.
\begin{enumerate}[align=left, labelwidth=4em,
leftmargin =\dimexpr\labelwidth+\labelsep\relax, font=$\bullet$~\normalfont\scshape]
\item[$S_0(\nu)$:] The input of the HEMT is terminated at ${T_\text{hot}\sim \SI{3.3}{K}}$.
\item[$S_1(\nu)$:] The input of the HEMT is terminated at ${T_\text{cold}\sim \SI{100}{mK}}$.
\item[$S_2(\nu)$:] The HEMT is connected to the parametric amplifier and MKID with the pump tone and DC current off.
\item[$S_3(\nu)$:] The HEMT is connected to the parametric amplifier and MKID with the pump tone and DC current on.
\end{enumerate}

The sum of the single sided power spectral densities for each quadrature around a frequency $f$, on a terminated  transmission line with impedance $Z_0$, and at an equilibrium temperature $T$ is given by
\begin{equation}\label{eq:termination_noise}
S(\nu) = 4 h f Z_0 \left(\frac{1}{e^{\sfrac{hf}{k_B T}} - 1} + \frac{1}{2} \right),
\end{equation}
where the first component comes from the thermal Johnson noise and the second from the zero point fluctuations.~\cite{Caves1982} Equation~\ref{eq:termination_noise} holds for $\nu \ll f$ and shows that the noise is independent of spectral frequency.

The noise added by the two different temperature terminations is calculated using equation~\ref{eq:termination_noise} and is labeled $S_\text{hot}(\nu)$ and $S_\text{cold}(\nu)$ for $T_\text{hot}$ and $T_\text{cold}$ respectively. The four measurements can then be written in terms of their components, where $S_\text{I}(\nu)$ is the noise at the input of the parametric amplifier, $S_\text{P}(\nu)$ is the noise added by the parametric amplifier, and $S_\text{H}(\nu)$ is the noise added by the HEMT amplifier.
\begin{equation} \label{eq:y_factor}
\begin{aligned}
  S_0(\nu) =& G_\text{H}(\nu)(S_\text{hot}(\nu) + S_\text{H}(\nu)) \\
  S_1(\nu) =& G_\text{H}(\nu)(S_\text{cold}(\nu) + S_\text{H}(\nu)) \\
  S_2(\nu) =& G_\text{H}(\nu) (S_\text{I}(\nu) + S_\text{H}(\nu)) \\
  S_3(\nu) =& G_\text{H}(\nu) (G_\text{P} (S_\text{I}(\nu) + S_\text{P}(\nu)) +  S_\text{H}(\nu)) \\
\end{aligned}
\end{equation}
$G_\text{H}(\nu)$ is the total gain of the system excluding the parametric amplifier and must be calculated. $G_\text{P}$ is the gain of the parametric amplifier and can be accurately determined by measuring the amplitude of a probe tone tuned off resonance with the parametric amplifier turned on and off. This system of equations can be solved for the unknown parameters.
\begin{equation} \label{eq:y_result}
\begin{aligned}
  G_\text{H}(\nu) =& \frac{S_0(\nu) - S_1(\nu)}{S_\text{hot}(\nu) - S_\text{cold}(\nu)}\\
  S_\text{H}(\nu) =& \frac{S_1(\nu) S_\text{hot}(\nu) - S_0(\nu) S_\text{cold}(\nu)}{S_0(\nu) - S_1(\nu)}\\
  S_\text{I}(\nu) =& \frac{(S_2(\nu) - S_1(\nu)) S_\text{hot}(\nu) + (S_0(\nu) - S_2(\nu)) S_\text{cold}(\nu)}{S_0(\nu) - S_1(\nu)}\\
  S_\text{P}(\nu) =& \frac{[S_3(\nu) - S_1(\nu) - G_\text{P}(S_2(\nu) - S_1(\nu))]S_\text{hot}(\nu) + [S_0(\nu) - S_3(\nu) - G_\text{P}(S_0(\nu) - S_2(\nu))] S_\text{cold}(\nu)}{G_\text{P} (S_0(\nu) - S_1(\nu))}
\end{aligned}
\end{equation}

The added noise in units of photon quanta for the components of the system are given in equation~\ref{eq:noise_numbers}, where we have defined $A_\text{sys}(\nu)$ to be the total noise of the system off resonance.
\begin{equation} \label{eq:noise_numbers}
\begin{aligned}
  A_\text{I}(\nu) =& \frac{S_\text{I}(\nu)}{4 h f Z_0} \\
  A_\text{P}(\nu) =& \frac{S_\text{P}(\nu)}{4 h f Z_0} \\
  A_\text{H}(\nu) =& \frac{S_\text{H}(\nu)}{4 h f Z_0} \\
  A_\text{sys}(\nu) \equiv& A_\text{I}(\nu) + A_\text{P}(\nu) + \frac{A_\text{H}(\nu)}{G_\text{P}}
\end{aligned}
\end{equation}

\section{Noise Measurement Statistical and Systematic Errors} \label{sec:systematics}
As written in appendix~\ref{sec:noise_calc}, $G_H(\nu)$, $S_\text{H}(\nu)$, $A_\text{H}(\nu)$, $S_\text{P}(\nu)$, $A_\text{P}(\nu)$, $S_\text{I}(\nu)$, $A_\text{I}(\nu)$, $A_\text{sys}(\nu)$ are all functions of $\nu$. However, they should be frequency independent. We can look for this characteristic to check for any unexpected properties of the data. It also allows us to easily determine the statistical uncertainties from the marginal likelihood distribution for the mean.
\begin{equation}
    \mathcal{L}\left(\bar X | X(\nu)\right) = \text{StudentT}\left(\frac{\bar X - E_\nu[X(\nu)]}{\sqrt{\frac{\Var_\nu[X(\nu)]}{N_\nu}}}, N_\nu - 1 \right),
\end{equation}
where $X(\nu)$ is one of the above parameters, $\E_\nu[\cdot]$ is the mean of the values over frequency, $\Var_\nu[\cdot]$ is the variance of the values over frequency, and $N_\nu$ is the number of frequency bins. For the amplifier added noise numbers, an additional prior can be included for only allowing physical values, $\bar X \ge \sfrac{1}{2}$.

We also account for any systematic errors which affect the noise measurement. Error terms are estimated and a Monte Carlo simulation is done to compute a posterior distribution for the noise numbers.

Losses at the \SI{100}{mK} stage are small but they may exist. We introduce two loss parameters. $l_\text{P}$ represents losses between the parametric amplifier and switch, and $l_\text{H}$ represents losses between the switch and HEMT. We expect these losses to be not much greater than \SI{1}{dB}, so we draw random samples from the following exponential distribution:
\begin{equation}
    f(x) = \begin{cases} \frac{1}{\beta - 1} e^{-\frac{x - 1}{\beta - 1}} & x \ge 1 \\ 0 & x < 1 \end{cases}
\end{equation}
with $\beta = \SI{1}{dB} \approx 1.26$.

We also consider thermometer calibration errors which we do not expect to be more than \SI{10}{\percent}. The temperatures for $S_\text{cold}(\nu)$ and $S_\text{hot}(\nu)$ are drawn from the normal distribution $\mathcal{N}\left(T, \sfrac{T}{10} \right)$ for their respective temperatures, where the second value is the standard deviation.

The resulting equations representing our measurement are modified from equation~\ref{eq:y_factor} to
\begin{equation}\label{eq:y_factor2}
\begin{aligned}
  S_0(\nu) =& \frac{G_\text{H}(\nu)}{l_\text{H}} (S'_\text{hot}(\nu) + S_\text{H}(\nu)) \\
  S_1(\nu) =& \frac{G_\text{H}(\nu)}{l_\text{H}} (S'_\text{cold}(\nu) + S_\text{H}(\nu)) \\
  S_2(\nu) =& \frac{G_\text{H}(\nu)}{l_\text{H}} \left(\frac{S_\text{I}(\nu)}{l_\text{P}} + S_\text{H}(\nu) \right) \\
  S_3(\nu) =& \frac{G_\text{H}(\nu)}{l_\text{H}} \left(\frac{G_\text{P}}{l_\text{P}} (S_\text{I}(\nu) + S_\text{P}(\nu)) +  S_\text{H}(\nu) \right). \\
\end{aligned}
\end{equation}
$S'_\text{hot}(\nu)$ and $S'_\text{cold}(\nu)$ are from the calibration terminations if their temperatures were different than our expectation. The solution to equation~\ref{eq:y_factor2} for many different samples of the loss and temperature distributions give the second set of confidence intervals presented in table~\ref{tab:noise}.

\section{Resonance Fitting} \label{sec:loop}
Each MKID is a notch resonator which has near perfect transmission far from the resonance frequency and a minimum in transmission near the resonance frequency. These resonances can be well modeled by an asymmetric Lorentzian function.
\begin{equation} \label{eq:loop}
    S_{21}(f) = \left( g_0 + g_1 f + g_2 f^2 \right) e^{i\phi_0-2 \pi i \tau f} \left(1 - \frac{Q \hat Q_c^{-1}}{1 + 2 i Q x} \right)
\end{equation}
The $g_0$, $g_1$, $g_2$, $\phi_0$, and $\tau$ terms represent a quadratic gain background, constant phase offset and cable delay for the system. The loop asymmetry is set by the complex coupling quality factor, $\hat Q_c$ and is related to the internal and total quality factors of the resonator using $Q^{-1} = Q_i^{-1} + Q_c^{-1} = Q_i^{-1} + \text{Re}\left[\hat Q_c^{-1} \right]$.~\cite{Khalil2012}

\begin{figure}[t]
\resizebox{\columnwidth}{!}{\input{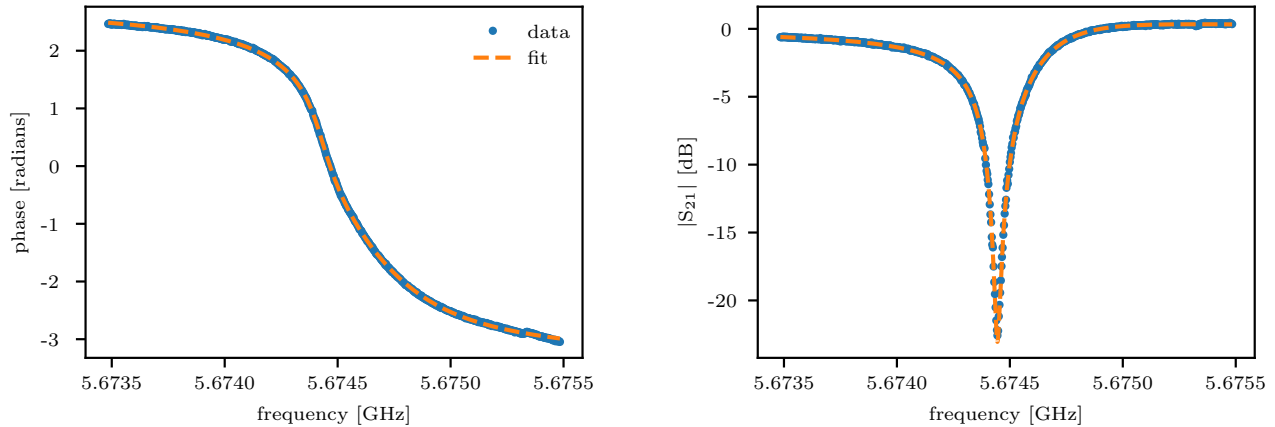}}
\caption{A fit to the phase (left) and magnitude (right) of the measured forward scattering parameter for the resonator is shown. The fitted gain baseline and cable delay are removed from the plot. The phase is referenced to the center of the resonance loop after the cable delay is removed.} \label{fig:loop}
\end{figure}

We use a nonlinear kinetic inductance model to describe the frequency detuning parameter, $x$, since the resonator described in this paper is driven close to its bifurcation power.~\cite{Swenson2013}
\begin{equation}
    Q x = Q \frac{f - f_r}{f_r} + \frac{a}{1 + 4 Q^2 x^2}
\end{equation}
The parameter, $a$, is linearly related to the generator power and ranges from 0 to about 0.77 before the onset of bifurcation. $f_r$ is the resonance frequency when $a = \num{0}$. A fit to the phase and magnitude of equation~\ref{eq:loop} is shown in figure~\ref{fig:loop}, which gives the parameter values $Q_i = \num{190000}$, $Q_c = \num{15100}$, and $a = 0.28$.

\section{Detector Response} \label{sec:response}
When an optical MKID is illuminated at photon count rates much lower than the characteristic decay rate of the sensor, we can measure individual photon energies and arrival times. The quasi-particle density resulting from a single-photon absorption can be measured by monitoring either the resonator's phase or dissipation quadratures. Each signal resembles that of a standard x-ray calorimeter response, and much of the formalism used to analyze this data is adapted from corresponding techniques generalized to this two-dimensional case.~\cite{Moseley1988, Szymkowiak1993, Fowler2017}

In an ideal detector for a fixed energy, the detector response should not vary. We model this response as an energy dependent pulse shape, $s_i(t, E)$, and amplitude, $A_i(E)$, where $i \in \{\theta, d\}$ corresponds to either the phase or dissipation response. The amplitude and shape are separated because for our detector the pulse shape is only weakly dependent on energy. $A_i(E)$ is also a good indicator of detector linearity. In practice, it is not possible to know these functions over a continuous range of energies, but they can be approximated by averaging many pulse records together for a discrete set of known photon energies. As long as the number of pulse records is large, this averaging procedure gives us noise free estimates of the detector response.

\begin{figure}[t]
\resizebox{\columnwidth}{!}{\input{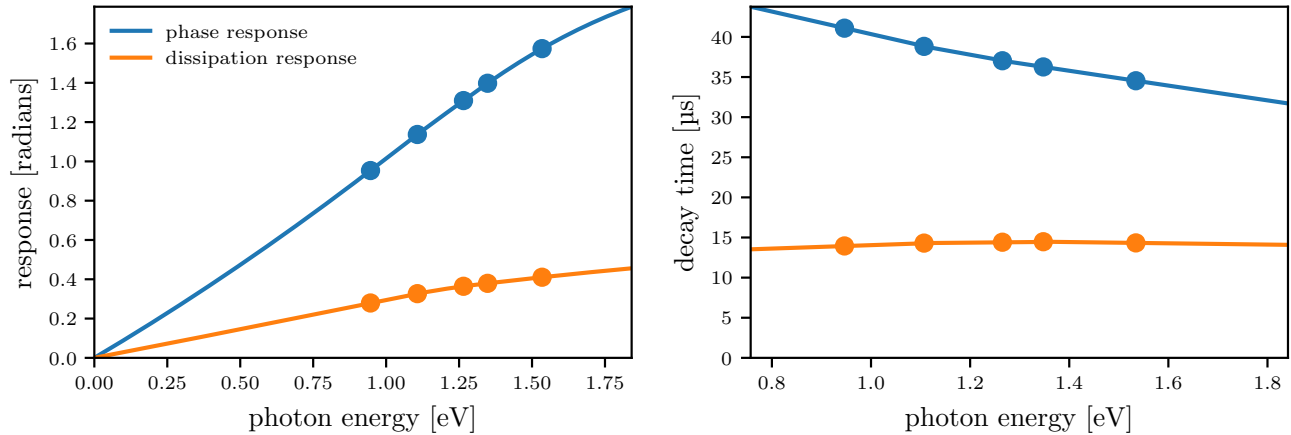}}
\caption{Left: The detector response as a function of energy is shown. Solid lines correspond to the interpolation constructed from the known energies (shown as points). These calibrations change slightly when the parametric amplifier is turned off because of the change in coupling quality factor. Right: The pulse shape decay time as a function of energy is shown. For the solid lines, the decay rate is computed using the interpolated pulse shape.} \label{fig:calibration}
\end{figure}

\begin{figure}[t]
\resizebox{\columnwidth}{!}{\input{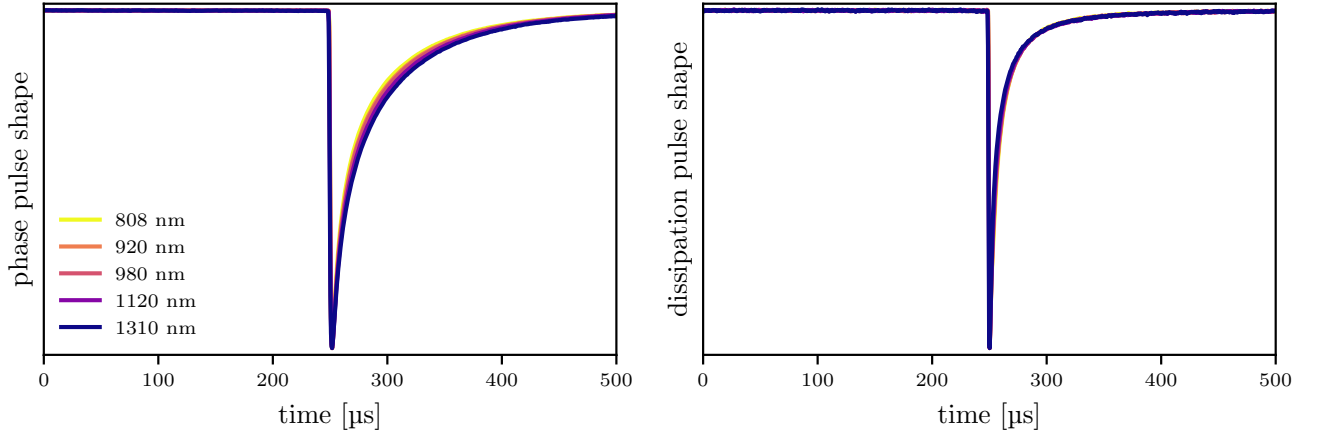}}
\caption{The pulse shapes for various photon energies are shown for the phase response (left) and dissipation response (right). Both shapes are normalized. The pulse shapes do not change when the parametric amplifier is turned off.} \label{fig:template}
\end{figure}

To determine the pulse shapes and amplitudes at all energies, we estimate the functions by interpolating between the known energies (five are used in our case). For the amplitudes, we use second order splines with ``not-a-knot'' boundary conditions, fixing the zero point. The amplitude calibration is shown in the left of figure~\ref{fig:calibration}. The detector is linear up to the highest energies tested where saturation effects begin to flatten the response. At these energies, the detector has the highest resolving power. For the pulse shapes, we linearly interpolate over each frequency in their Fourier transforms. We find that this procedure preserves the pulse shape normalization to a very good approximation as long as enough known energies are used. The pulse decay time is defined as the integrated area under the normalized pulse shape and is shown as a function of energy on the right of figure~\ref{fig:calibration}. This definition is adopted for it's simplicity and because the pulse shapes do not fit well to an exponential function. The pulse shapes for each energy are plotted in figure~\ref{fig:template}. Both of these interpolations introduce some systematic calibration error for photon measurements of significantly different energy than that of the calibration points. In this paper, however, the measured energies are the same as the calibration energies, so this effect is negligible.

The pulse shapes change very little except for the decay time in the phase quadrature which decreases with energy. This effect might be explained by a higher quasi-particle recombination rate at higher quasi-particle densities. However, the same tendency is not seen in the dissipation quadrature. The decay time in the dissipation quadrature is additionally much lower than in the phase quadrature. This behavior is a characteristic of the superconducting PtSi from which the resonator patterned. Similar results have been seen in TiN superconducting resonators~\cite{Gao2012} and may be attributable to a small population of sub-gap states.

\section{Pulse Fitting} \label{sec:fitting}
After the detector response  amplitude and shape has been accurately measured, determining an absorbed photon's energy and arrival time becomes an optimization problem. We model the pulse records with the following equation:
\begin{equation} \label{eq:model}
    \boldsymbol{m}(t; E, t_0) = \begin{pmatrix} A_\theta(E) \; s_\theta(t - t_0, E) \\  A_d(E) \; s_d(t - t_0, E) \end{pmatrix}.
\end{equation}
Equation~\ref{eq:model} represents a two-dimensional time dependent vector that we expect to measure for a given photon energy, $E$, and arrival time, $t_0$. The actual data contains noise, so we compute a maximum likelihood estimate for the photon energy and arrival time by minimizing a $\chi^2$ function with respect to $t_0$ and $E$. For a given data record, $\boldsymbol{d}(t)$,
\begin{equation} \label{eq:chi2}
    \chi^2 = \int_{-\infty}^\infty df (\boldsymbol{\tilde d} - \boldsymbol{\tilde m})^\dag \boldsymbol{S}^{-1} (\boldsymbol{\tilde d} - \boldsymbol{\tilde m}),
\end{equation}
where a tilde denotes a Fourier transform with respect to time and $\boldsymbol{S}$ is the spectral noise covariance matrix determined by analyzing data with no photon events.

We compute $\chi^2$ in the frequency domain to simplify the computation, and because by not including the zero frequency in the integration, any small DC offset in the data is ignored. Correcting for the DC offset is important since our detectors have significant low-frequency noise below the measurement bandwidth. Because $\boldsymbol{m}(t; E, t_0)$ is a nonlinear function of it's parameters, no closed form solution can be used to find the best estimates of $E$ and $t_0$. Instead, we use a standard least-squares minimization routine to compute the results.

This formalism assumes following:
\begin{enumerate}
\item The photon records are completely isolated in time.
\item Each photon of the same energy creates the same detector response up to an additive noise.
\item The added noise is Gaussian and stationary (i.e. does not change as quasi-particles are broken in the detector).
\end{enumerate}
When all of these assumptions are satisfied, the expected variance of each estimator can be calculated from equation~\ref{eq:chi2} giving
\begin{equation} \label{eq:variance}
\begin{aligned}
    \sigma_\alpha^2 =& \left. \left[ \frac{1}{2} \frac{\partial^2 \chi^2}{\partial \alpha^2} \right]^{-1} \right|_{\hat \alpha} \\
    =& \left. \left[ \int_{-\infty}^\infty df \; (\partial_\alpha \boldsymbol{\tilde m})^\dag \; \boldsymbol{S}^{-1} \; (\partial_\alpha \boldsymbol{\tilde m}) \right]^{-1} \right|_{\hat \alpha}, \\
\end{aligned}
\end{equation}
where $\alpha \in \{E, t_0\}$, $\hat \alpha$ represents the maximum likelihood value, and $\partial_\alpha$ represents a partial derivative with respect to the estimated parameter.

Equation~\ref{eq:variance} is a good estimate of the expected variance of the energy and arrival time model parameters only if the above assumptions are satisfied. We address the first assumption by ensuring that the photon absorption count rate is much slower than the characteristic decay rate of the detector and by excluding photons with nearby arrival times from the analysis. The second assumption is a detector property and may or may not be true depending on the detector design. The third assumption, however, is more difficult to assess. While the noise is very nearly Gaussian, the detector's phase noise decreases as the resonance frequency moves off of the signal tone during a photon absorption event. Adequately accounting for this effect is still a debated problem and beyond the scope of this paper.~\cite{Fixsen2014, Shank2014, Fixsen2004} However, since the phase noise is smaller at the peak of the pulse record, Monte Carlo simulations of our data, presented in appendix~\ref{sec:mc}, suggest that equation~\ref{eq:variance} is a lower bound on the true variance of the estimate as long as assumption 2 is satisfied. This intuitively makes sense, since the average noise in the pulse record is less than that used to make the estimation.

\section{Pulse Measurement Baseline} \label{sec:baseline}
\begin{figure}[t]
\resizebox{\columnwidth}{!}{\input{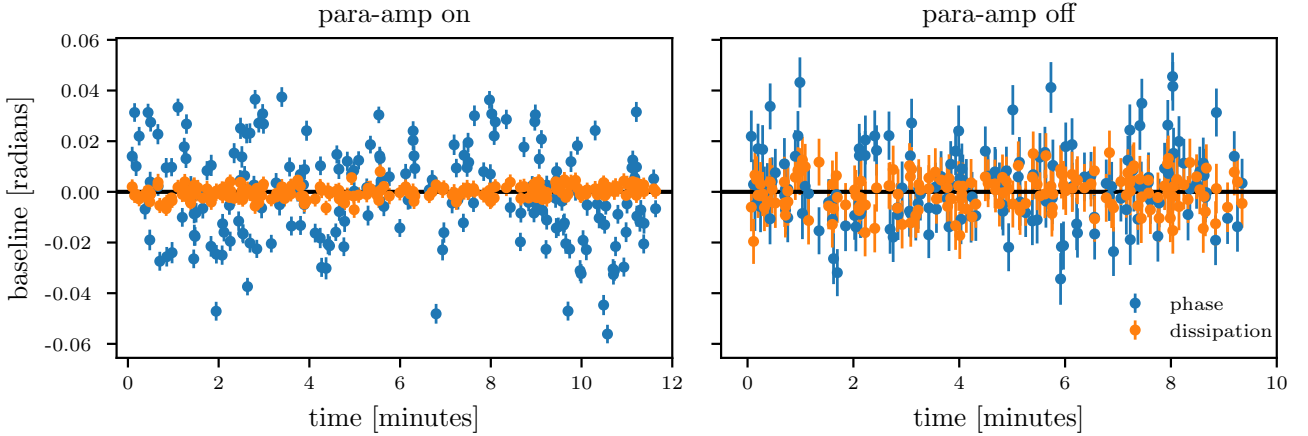}}
\caption{The measured pre-trigger baseline in each signal quadrature is shown as a function of time with the parametric amplifier on and off. Error bars correspond to statistical \SI{95}{\percent} confidence intervals. A random subset of the data in this time interval is plotted so that the error bars can be seen. The baseline is stable to within the error over the many minutes needed to take a single laser measurement.} \label{fig:baseline}
\end{figure}

Noise below the measurement bandwidth contributes a small DC offset to each recorded photon event. By measuring the pre-trigger baseline of each pulse, we can see, in figure~\ref{fig:baseline}, how this baseline varies over the measurement duration. On the \SI{\sim10}{\minute} time scale, the baseline stays constant to within the measurement error.

The spread of measured baselines in the phase quadrature does not change much when the parametric amplifier is turned on because the low frequency noise is dominated by the two-level system noise in the detector. The two-level system noise corresponds to real resonance frequency shifts which could change the pulse height or shape through the nonlinear phase to frequency relationship shown in figure~\ref{fig:loop}. We will find it useful to decompose the baseline into these two effects in order to simulate our detector. The baselines can be modeled as normal random variables:

\begin{equation}
\begin{aligned}
    \delta_\theta &= \delta_\text{TLS} + \delta_\text{sys} \\
    \delta_d &= \delta_\text{sys}, \\
\end{aligned}
\end{equation}
where the variances are related by the usual equation,
\begin{equation}
    \sigma_{\delta_\theta}^2 = \sigma_{\delta_\text{TLS}}^2 + \sigma_{\delta_\text{sys}}^2.
\end{equation}

\section{Pulse Fitting Monte Carlo Simulations} \label{sec:mc}
To be confident in the results of equation~\ref{eq:variance} and in our fitting routine, we simulate our data assuming a uniform pulse height for each energy and compute the resulting fitted energies. Our fake data follows the general form
\begin{equation} \label{eq:mc}
    \boldsymbol{d}(t; E) = \begin{pmatrix} A_\theta(E) \; s_\theta(t - \delta_{t_0}, E) + \delta_\theta \\  A_d(E) \; s_d(t - \delta_{t_0}, E)  + \delta_d \end{pmatrix} + \boldsymbol{n}(t).
\end{equation}
Our pulse trigger is accurate to the level of the sample spacing, \SI{0.5}{\micro \second}, but there is some jitter. We model this effect with the random variable $\delta_{t_0}$ drawn from a normal distribution with a standard deviation of \SI{0.25}{\micro \second}. A covariance matrix for the noise is estimated using the noise data taken directly before the measurement. $\boldsymbol{n}(t)$ is then drawn from the multivariate normal distribution with the corresponding covariance.

As discussed in appendix~\ref{sec:baseline}, baseline drifts due to two-level systems could cause extra variations in the pulse height and shape. We cannot model this effect perfectly because the phase and dissipation trajectory of our pulses is not well understood. However, if we ignore the dissipation response, we can estimate the magnitude of this effect using $\theta(f)$ and $f(\theta)$ from figure~\ref{fig:loop}. Equation~\ref{eq:mc} becomes
\begin{equation} \label{eq:mc_baseline}
    \boldsymbol{d}(t; E) =
    \begin{pmatrix} \begin{aligned}
\theta(f(&A_\theta(E) \; s_\theta(t - \delta_{t_0}, E)) + f(\delta_\text{TLS})) \!\!\!\!\! &+\delta_\text{sys} \\ &A_d(E) \; s_d(t - \delta_{t_0}, E) &+ \delta_\text{sys}
\end{aligned} \end{pmatrix} + \boldsymbol{n}(t).
\end{equation}

As a final check, we can also account for the non-stationary noise in our simulated pulse records. We use the \num{2000} pulse records with estimated energies closest to the laser energy and have the same arrival time as the model. The noise is recovered by subtracting off the model from the pulse records, and a covariance matrix is estimated similarly to before. This procedure produces noise that looks realistic, but we expect there to be problems near the rise of the pulse. Small variations in the arrival time between the model and data can cause significant errors in the subtraction in this region. For this reason and because the exact way to account for baseline effects is unknown, the Monte Carlo simulations are not used for the expected values in table~\ref{tab:R}. The results are included only as a check that the non-stationary noise does not degrade the energy resolution.

\begin{figure}[t]
\resizebox{\columnwidth}{!}{\input{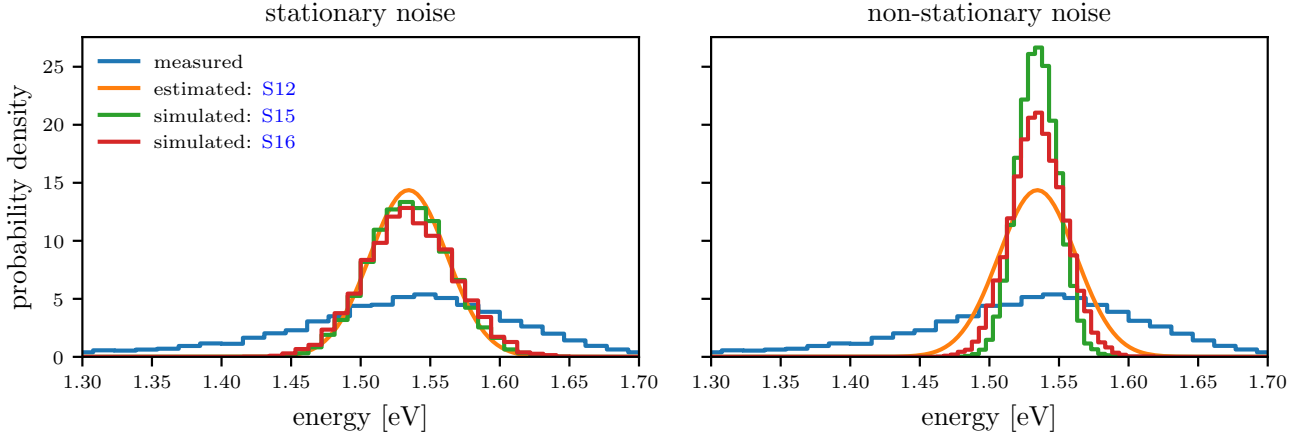}}
\caption{Plotted are Monte Carlo simulation results with different data models for \SI{808}{nm} photons alongside actual data, $R \sim \num{8.9}$, and the expected distribution from equation~\ref{eq:variance}, $R \sim \num{23}$. The simulated resolving powers using equations~\ref{eq:mc} and~\ref{eq:mc_baseline} are \num{22}, \num{20} (left) and \num{38}, \num{32} (right) respectively.} \label{fig:mc}
\end{figure}

At \SI{808}{nm} the device calibration is the most nonlinear, so in figure~\ref{fig:mc} we present the Monte Carlo results for this energy using the simulated data described above. We find good agreement between equation~\ref{eq:variance} and the simulations for the stationary noise case. Baseline drifts slightly degrade the resolving power, but the decrease is mostly negligible. For non-stationary noise, we recover a higher resolving power as expected, and two-level system baseline effects significantly reduce the simulated resolving power. These results support the claim that the resolving power given by equation~\ref{eq:variance} represents a lower bound on the expected resolving power for this data.

\section{Current Distribution} \label{sec:current}
The data presented in table~\ref{tab:R} show a discrepancy between the measured and expected resolving powers. However, the pulse fitting algorithm performs as expected on simulated data computed using the enumerated assumptions discussed in appendix~\ref{sec:fitting}. The discrepancy between the real and simulated data, therefore, must be either in the noise model or response model for the detector. Here we investigate one possible explanation for the detector response not being fixed for a given photon energy.

\begin{figure}[b]
\resizebox{\columnwidth}{!}{\input{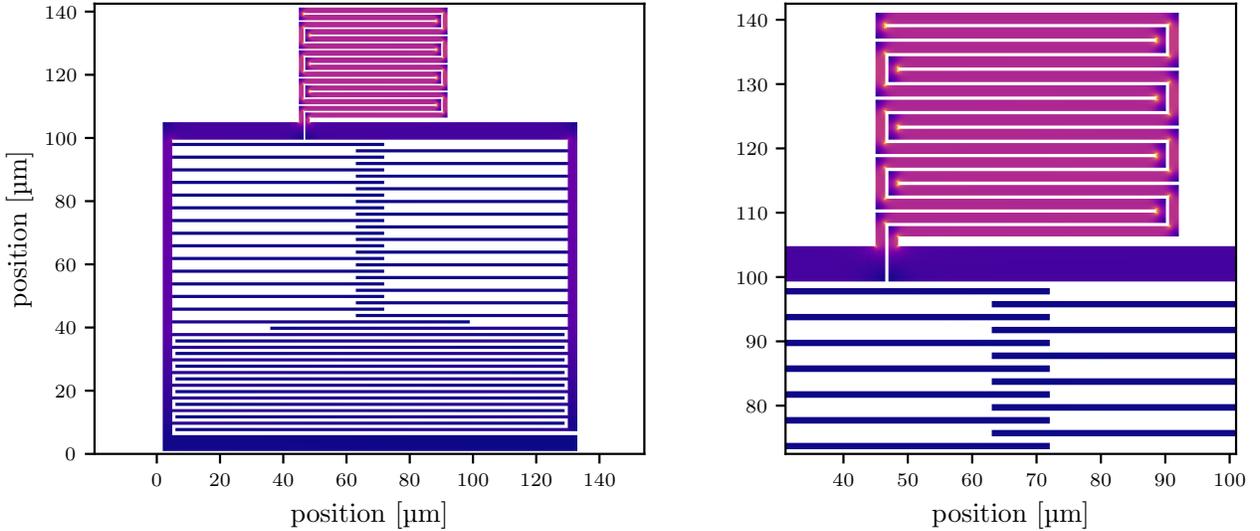}}
\caption{The simulated current density for the full resonator used in this paper is shown on the left. The photon-sensitive portion of the resonator is shown in more detail on the right. The scale for the current density is arbitrary.} \label{fig:current}
\end{figure}

The response of a MKID induced by a uniform distribution of quasiparticles differs from the response from a localized quasiparticle distribution. In the localized case, the detector response is proportional to the square of the current density multiplied by the quasiparticle distribution integrated over each position in the resonator.~\cite{Gao2008} In an ideal lumped element resonator, all of the current is in the inductor and none is in the capacitor. When designing these detectors, then, it is important to ensure a uniform current distribution in the inductor.

The current distribution was computed in a Sonnet simulation for the resonator used in this paper and is shown in figure~\ref{fig:current}. It takes the resonator about \SI{1}{\micro s} to respond to a photon absorption event, and in that time, the quasiparticles will have diffused an amount determined by the superconductor's diffusion constant. This time scale roughly sets the amplitude of a photon pulse since at longer time scales the quasiparticles begin to recombine (see figure~\ref{fig:template}). In similar films the diffusion constant has been measured to be \SIrange{\sim2}{8}{cm^2/s},~\cite{Baturina2005} so we use this range for our analysis.

\begin{figure}[t]
\resizebox{\columnwidth}{!}{\input{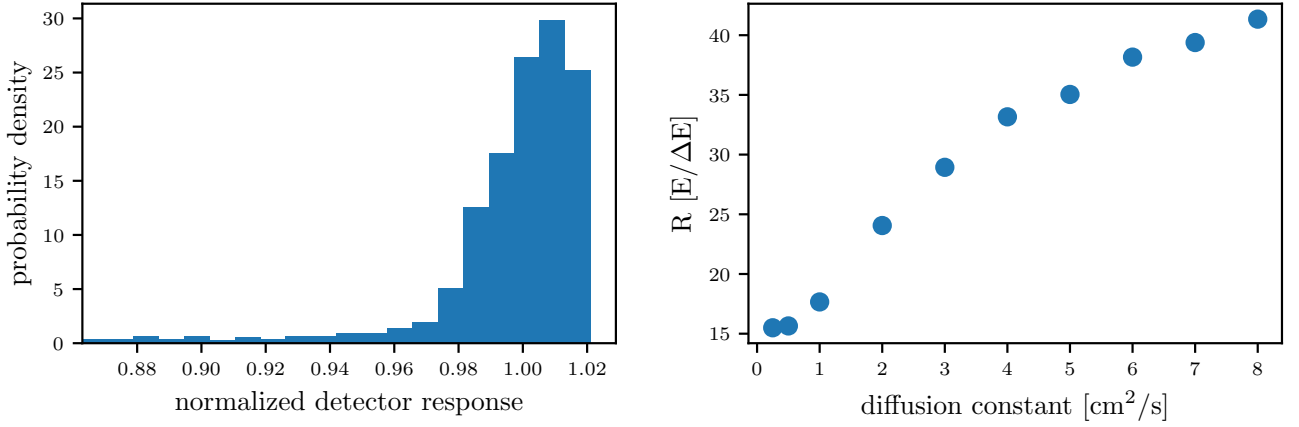}}
\caption{The effects of the current non-uniformity on the detector resolving power are shown. All other sources of noise are ignored. For a random photon absorption location on the inductor, the distribution of responses is shown for a diffusion constant of \SI{3}{cm^2/s} (left). Larger diffusion constants result in higher resolving powers (right).} \label{fig:R}
\end{figure}

To model the effect of the current non-uniformity, we randomly selected photon absorption locations on the inductor, solved the diffusion equation for each one, and computed the resulting response. Photon strikes in the capacitor have normalized responses \num{\ll0.5} and are excluded from the calculation to increase the throughput. The distribution of responses for several different diffusion constants was computed to estimate its effect on the resolving power. The results are shown in figure~\ref{fig:R}.

\section{Hot Phonon Escape} \label{sec:phonon}
After a photon is absorbed by the superconducting PtSi in our detector, the resulting quasiparticle distributes its energy among more broken Cooper pairs and phonons. This energetic cascade ultimately results in an equilibrium distribution of quasiparticles on a roughly \SI{1}{ns} timescales.~\cite{Kozorezov2007} Phonons created in the intermediary stages of the cascade can escape the superconductor to the substrate, effectively removing energy from the measurement. The statistics of this process offer another potential explanation for the resolving power discrepancy presented in table~\ref{tab:R}.

The resolving power of our detector absorbing a photon with energy, $E$,  is given by
\begin{equation}\label{eq:R}
    R \equiv \frac{E}{\Delta E} = \frac{\num{1}}{\num{2.355}} \sqrt{\frac{E}{(F + J) \epsilon}},
\end{equation}
considering only Fano fluctuations and fluctuations in energy loss due to the escape of athermal phonons. $\epsilon$ is the mean energy required to generate one quasiparticle in the superconductor.~\cite{Kozorezov2007} $F$ is the Fano factor, typically assumed to be about 0.4 for most materials.~\cite{Fano1947} The $J$ factor accounts for the hot phonon escape and can be calculated using~\cite{[][{. Note: equation~\ref{eq:J} contains an extra factor of 4 which was missing from the original manuscript. This was determined through private communication with the authors.}]Kozorezov2008}
\begin{equation} \label{eq:J}
    J = \eta_t \frac{\Omega_D}{\epsilon} \frac{l_{ph}}{d} \frac{\num{12} (\num{1} + \lambda)}{\num{11} (\num{1}+\lambda) + \num{3}} \: g_1\!\!\left(\frac{\Omega_D}{\Omega_1}\right).
\end{equation}
$\Omega_D$ and $\Omega_1$ are the Debye energy and and the lower boundary energy for the phonon controlled down-conversion. $l_{ph}$ is the  mean free path for a phonon at the Debye frequency. $d$ is the film thickness. $\lambda$ is the effective electron-phonon coupling constant. $g_1(\cdot)$ is a dimensionless function defined by Kozorezov et al.

\bgroup
\def\arraystretch{1.5}
\ctable[
caption={Parameters used in the evaluation of equation~\ref{eq:R} are tabulated here. The top half corresponds to known or estimated values. Parameters computed from those given above are listed on the bottom half.},
label={tab:parameters},
width=\columnwidth,
pos=t,
botcap
]{
XXp{2.8in}l
}{
}{
\toprule[0.08em]  
Parameter & Value & Description & Comment \\
\cmidrule[0.03em]{1-4}
F & \num{0.4} & Fano factor & typical value for superconductors~\cite{Fano1947}\\
$T_c$ & \SI{930}{mK} & superconducting transition temperature & typical value for a MEC array~\cite{Szypryt2017b}\\
$N_0$ & \SI{1}{state\per eV\per ion} & density of states at the Fermi energy & theoretical~\cite{Koc2011} \\
$v_s$ & \SI{3600}{m\per s} & speed of sound in PtSi & theoretical~\cite{Koc2011} \\
$\Omega_D$ & \SI{31}{meV} & Debye energy & theoretical~\cite{Koc2011} \\
$\left< \alpha^2 \right>_\text{avg}$ & \SI{1.5}{meV} & averaged squared electron-phonon interaction  & typical value for superconductors~\cite{Kaplan1976} \\
$\mu^*$ & \num{0.13} & Coulomb pseudo-potential & typical value for superconductors~\cite{McMillan1968} \\
$\eta_t$ & 1 & phonon transmission coefficient & assumed for PtSi-sapphire boundary \\
$d$ & \SI{50}{nm} & PtSi film thickness & typical value for a MEC array~\cite{Szypryt2017b} \\
\cmidrule[0.03em]{1-4}
$\Delta_0$ & \SI{0.14}{meV} & zero temperature superconducting gap energy & $\num{1.764}\; k_B \; T_c$\\
$\epsilon$ & \SI{0.25}{meV} & mean quasiparticle creation energy & $\num{\sim1.75} \; \Delta_0$~\cite{Kozorezov2000} \\
$l_{ph}$ & \SI{2.0}{nm} & phonon mean free path at $\Omega_D$ & equation~\ref{eq:mfp} \\
$\lambda$ & \num{0.43} & effective electron-phonon coupling constant & equation~\ref{eq:lambda} \\
$\Omega_1$ & \SI{5.3}{meV} & threshold energy for the phonon cascade & equation~\ref{eq:omega1}\\
$g_1\!\left(\sfrac{\Omega_D}{\Omega_1}\right)$ & \num{0.55} & special function & numerically evaluated~\cite{Kozorezov2008} \\
$J$ & \num{2.5} & phonon noise factor & equation~\ref{eq:J} \\
\bottomrule[0.08em]
}
\egroup

$\eta_t$ is the phonon transmission coefficient from PtSi to sapphire. While sapphire is significantly harder than PtSi, PtSi is more dense. This relationship leads to the acoustic impedance of the two materials being fairly well matched, so we will take $\eta_t \sim 1$.

We can write some of the parameters in equation~\ref{eq:J} in terms of more tractable material parameters.~\cite{Kaplan1976, Kozorezov2008}
\begin{equation} \label{eq:mfp}
    l_{ph} = \frac{\hbar v_s}{\num{8} \pi N_0 \left< \alpha^2 \right>_\text{avg} \Omega_D},
\end{equation}
and
\begin{equation} \label{eq:omega1}
    \Omega_1 = \Omega_D \sqrt{\frac{\num{2}}{\num{3}} (\num{1} + \lambda)N_0 \Omega_D},
\end{equation}
where $N_0$ is the single spin density of states at the Fermi energy, $v_s$ is the speed of sound in the superconductor, and $\left< \alpha^2 \right>_\text{avg}$ is the averaged squared electron-phonon interaction energy.

For the case of PtSi, we don't have a good measurement of $\lambda$, but it can be estimated using~\cite{McMillan1968}
\begin{equation} \label{eq:lambda}
    \lambda = \frac{\num{1.04} + \mu^* \ln\left(\sfrac{\Omega_D}{\num{1.45} k_B T_c} \right)}{(1 - \num{0.62} \mu^*)\ln\left(\sfrac{\Omega_D}{\num{1.45} k_B T_c} \right) - \num{1.04}}.
\end{equation}
$T_c$ is the superconducting transition temperature. $\mu^*$ is the Coulomb pseudo-potential which is also unknown, but generally takes a very small range of values for different materials. We make the approximation $\mu^*\sim \num{0.13}$.

The values used to compute the resolving power from equation~\ref{eq:R} are listed in table~\ref{tab:parameters}, resulting in $R\sim\num{20}$ for \SI{808}{nm} photons and $R\sim\num{15}$ for \SI{1310}{nm} photons. While a few of the parameters needed for this calculation are well known for our array, some had to be taken from theoretical calculations and could be inaccurate. Moreover, the values of $\epsilon$, $F$, $\mu^*$, and $\left< \alpha^2 \right>_\text{avg}$ are completely unknown for PtSi and had to be estimated from values for other superconductors. The largest error likely comes from our choice in $\left< \alpha^2 \right>_\text{avg}$. Considering the extremes of the values presented by Kaplan et al., we expand these estimates to $R \sim \numrange{13}{30}$ for \SI{808}{nm} photons and $R \sim \numrange{10}{24}$ for \SI{1310}{nm} photons

\end{document}